\documentclass[reprint,aps,showpacs,pra]{revtex4-1}
  \usepackage{color}
  \usepackage{newlfont}
  \usepackage{graphicx}
  \usepackage{amssymb}
  \usepackage{amsmath}
  \usepackage{bm}
  \usepackage{verbatim}
  \usepackage[latin1]{inputenc}
  \usepackage{bbm}
  \usepackage{multirow}
  \usepackage[thinlines]{easytable}
  \usepackage{stackrel}
 \usepackage[varg]{txfonts}

\newcommand{\ket}[1]{\mbox{$ | #1 \rangle $}}
\newcommand{\bra}[1]{\mbox{$ \langle #1 | $}}
\newcommand{\be}{\begin{equation}}
\newcommand{\ee}{\end{equation}}
\newcommand{\beq}{\begin{eqnarray}}
\newcommand{\eeq}{\end{eqnarray}}

\begin{document}

\title{Information-reality complementarity: The role of measurements and quantum reference frames}
\author{P. R. Dieguez}
\author{R. M. Angelo}
\affiliation{Department of Physics, Federal University of Paran\'a, P.O. Box 19044, 81531-980, Curitiba, Paran\'a, Brazil}

\begin{abstract} 
Recently, a measure has been put forward which allows for the quantification of the degree of reality of an observable for a given preparation [A. L. O. Bilobran and R. M. Angelo, Europhys. Lett. {\bf 112}, 40005 (2015)]. Here we employ this quantifier to establish, on formal grounds, relations among the concepts of measurement, information, and physical reality. After introducing mathematical objects that unify weak and projective measurements, we study scenarios showing that an arbitrary-intensity unrevealed measurement of a given observable generally leads to an increase of its reality and also of its incompatible observables. We derive a complementarity relation connecting an amount of information associated with the apparatus with the degree of irreality of the monitored observable. Specifically for pure states, we show that the entanglement with the apparatus precisely determines the amount by which the reality of the monitored observable increases. We also point out some mechanisms whereby the irreality of an observable can be generated. Finally, using the aforementioned tools, we construct a consistent picture to address the measurement problem.
\end{abstract}


\maketitle

\section{Introduction}

At every instant of time we probe our surroundings through a huge number of sequential projective measurements which induce us to believe that everything is real. When, for instance, we look at an object at rest on the ground, our eyes collect a bunch of photons which bring us information about the object. Because macroscopic objects are only slightly disturbed by the scattered photons, such measurements can be repeated many times yielding always the same information about the object. This process along with the ubiquitous verification of Newtonian behavior for macroscopic objects feed the illusion of a Laplacian determinism, according to which the physical properties of systems are well defined at all instants of time regardless of any external monitoring. Equipped with the superposition principle, quantum mechanics teaches us, however, that this view cannot be generally maintained. In fact, it has repeatedly been shown by experiments with isolated microscopic systems that the classical notion of reality is objectionable.

This apparent conflict between our fundamental theory of nature and the preconception of an observer-independent reality has always bothered the physical community and it seems fair to say that it remains as one of the most intriguing problems of quantum mechanics. Among the historical approaches to the issue, the criticism raised by Einstein, Podolsky, and Rosen (EPR) against quantum theory \cite{EPR35} caused a particularly great impact. Under the premise of locality, EPR argued that incompatible observables could be simultaneously real in scenarios involving entangled states. Since quantum mechanics is not able to simultaneously describe such elements of reality, it presumably is, according to EPR, an incomplete theory. However, as has been seminally pointed out by Bell \cite{bell64} and recently confirmed by loophole-free experiments \cite{hensen15,giustina15,shalm15}, the correlations observed in isolated microscopic systems cannot be described by theories supplemented with local-causal hidden variables. On the other hand, as Bohm has shown by explicit construction~\cite{bohm52}, it is perfectly viable to have a realistic hidden-variable theory, but at the expense of local causality.

In recent decades, conceptual advances concerning the emergence of objective reality from the quantum substratum have been obtained by use of mechanisms such as weak measurements~\cite{vaidman96}, decoherence~\cite{schlosshauer05}, and quantum Darwinism~\cite{zurek09,horodecki15}. Impacting results have also been reported about the ontology of the wave function~\cite{pusey12,lewis12,colbeck12,hardy13,patra13,aaronson13,leifer14,barrett14,branciard14,ringbauer15}. More recently, Bilobran and Angelo (BA) put forward an operational scheme to assess elements of reality \cite{bilobran15}. In a protocol involving preparation, unrevealed measurements, and quantum state tomography, they introduced a quantifier for the degree of irreality of an observable for a given state preparation. Among its many interesting properties, this measure has proven relevant in scenarios involving coherence \cite{angelo15} and nonlocality~\cite{gomes17}.
 
Despite all these efforts towards a profound understanding of the physical reality, too little (if any) has been achieved with regard to formal connections between elements of reality and fundamental concepts such as information and quantum correlations. The situation is no better when we try to understand the emergence of reality from the measurement process, which is a major conundrum of quantum theory. Contributing to filling this gap is the goal of this work. In contexts involving measurements of generic intensity, with outcomes revealed or not, we aim at deriving formal relations between BA's irreality and quantifiers of information, such as the mutual information and the von Neumann entropy. In particular, we want to learn what type of physical mechanisms can produce alterations in the degree of reality of observables and also shed some light on the drama originally proposed by Everett concerning a quantum measurement as seen from the perspective of two distinct observers \cite{everett57}.

This paper is structured as follows. Section~\ref{preliminaries} starts with a review of BA's measure of irreality and of some well-known objects of quantum information theory. In Secs.~\ref{measurement} and \ref{monitoring} we then introduce, as our first contribution, a map that conveniently interpolates between a weak and a projective measurement and a second map, defining a procedure that we call monitoring, that extends the first one to the context of unrevealed measurements. In Sec.~\ref{results} we present our main contributions. We show that dynamics involving arbitrary-intensity interactions and some type of discard invariably lead to an increase of reality. On the other hand, we find that local irreality can be created through both revealed measurements of arbitrary intensities and unitary dynamics marked by an effective violation of some conservation law. Remarkably, we derive a complementarity relation between the information acquired by the detection system and the degree of irreality of the probed observable. Finally, using this information-reality duality, we move to the two-observer drama proposed by Everett to discuss several aspects of the measurement problem, including the objectivity of reality, the classicality of the apparatus, the role of the reference frame, and the irreversibility of the measurement. Our approach addresses the measurement problem without invoking an external reservoir, that is, it uses only internal mechanisms of decoherence. We then close this work in Sec.~\ref{conclusion} with our concluding remarks.

\section{Preliminaries}
\label{preliminaries}

\subsection{Elements of reality}
\label{irreality}

There is not a unique view of physical reality, but it seems that in all of them the notion is related to the definiteness of physical quantities. In their celebrated work \cite{EPR35}, EPR introduce what they call a sufficient condition for the existence of an element of reality: ``If, without disturbing the system in any way, we can predict with certainty (that is, with a probability equal to unity) the value of a physical quantity, then there is an element of physical reality corresponding to this physical quantity.'' For uncorrelated systems, this criterion makes clear reference to eigenstates.

In his reply to EPR, Bohr~\cite{bohr35} argues in terms of his complementarity principle, according to which the elements of reality of incompatible observables cannot be established in the same experiment, but only through mutually excluding experimental arrangements. Thus, one cannot claim simultaneous reality for incompatible observables within the same experimental instance, even when entangled states are involved. In addition, for Bohr one cannot speak of the nature of microscopic systems before making a measurement. This perspective refutes EPR's rationale and elects the correlations generated in the experimental setup as the mechanism responsible for the establishment of physical reality (see Refs.~\cite{bilobran15,angelo15} for related discussions). In the same year, Ruark pointed out that EPR's conclusion derived from the adoption of a criterion that ``is directly opposed to the view held by many theoreticians, that a physical property of a given system has reality only when it is actually measured''~\cite{ruark35}.

Inspired by EPR's criterion, Redhead proposes \cite{redhead87}: ``If we can predict with certainty, or at any rate with probability one, the result of measuring a physical quantity at time $t$, then, at time $t$, there exists an element of reality corresponding to this physical quantity and having value equal to the predicted measurement result.'' Although apparently similar to EPR's definition, this one is intended to soften the condition on the relativistic causality hypothesis.

Realizing that a point common to all of these definitions is the relation with actual results of quantum measurements, Vaidman then proposes that ``for any definite result of a measurement there is a corresponding element of reality''~\cite{vaidman96}. Regarding ``definitive result'' as the definite shift of the probability distribution of the pointer variable, he suggests the following definition of elements of reality: ``If we are certain that a procedure for measuring a certain variable will lead to a definite shift of the unchanged probability distribution of the pointer, then there is an element of reality: the variable equal to this shift.'' With that, Vaidman extends the discussion of physical reality to the context of weak measurements.

Other works have argued that a better understanding of the physical nature can be achieved through the concept of information. Bruckner and Zeilinger defend that quantum physics is an elementary theory of information~\cite{brukner99,brukner03} and that even though information should not be taken as replacing the notion of reality, in their approach ``the notions of reality and of information are on equal footing''~\cite{zeilinger99}, which suggests some ontological status for information. Quantum Bayesianism, on the other hand, is a reconstruction of quantum mechanics that mixes subjective elements, associated with the probabilistic information that an agent has about the world, with objective elements, which are identified as the Hilbert space dimension of the quantum systems: ``Dimension is something a body holds by itself, regardless of what an agent thinks of it''~\cite{fuchs16}. For a recent overview of conceptions of reality in physics we refer the reader to Ref.~\cite{krizek17}.

Throughout the present paper, we employ a notion of reality that has recently been introduced by BA~\cite{bilobran15}. Its main advantage is that it is quantitative and operational. Bilobran and Angelo consider a preparation $\rho\in\cal{H_A\otimes H_B}$ submitted to a protocol of unrevealed measurements of a generic observable $A=\sum_aaA_a$, with projectors $A_a=\ket{a}\bra{a}$, acting on $\cal{H_A}$. Since the outcome of the measurement is kept secret, the resulting state reads
\be \label{PhiA}
\Phi_A(\rho):=\sum_a\left(A_a\otimes\mathbbm{1}_\cal{B}\right)\,\rho\,\left(A_a\otimes\mathbbm{1}_\cal{B}\right)=\sum_ap_aA_a\otimes\rho_{\cal{B}|a},
\ee 
where $\rho_{\cal{B}|a}=\bra{a}\rho\ket{a}/p_a$ and $p_a=\text{Tr}[(A_a\otimes\mathbbm{1}_\cal{B})\rho]$.
Under the premise that a measurement establishes the reality of an observable, BA propose to take $\Phi_A(\rho)$ as a state of reality for $A$ and $\rho=\Phi_A(\rho)$ as a formal criterion of reality. With that we can compute the degree of irreality of the observable $A$ given the preparation $\rho$ as 
\be \label{frakI}
\mathfrak{I}(A|\rho):=S(\Phi_A(\rho))-S(\rho),
\ee 
where $S(\rho)=-\text{Tr}(\rho\ln{\rho})$ stands for the von Neumann entropy. The above formula can be viewed as an entropic distance between the state $\rho$ under scrutiny and the state of reality $\Phi_A(\rho)$. This quantifier is non-negative and vanishes if and only if $\rho=\Phi_A(\rho)$. Also, it can be shown that the following decomposition holds:
\be \label{LIQC}
\mathfrak{I}(A|\rho)=\mathfrak{I}(A|\rho_{\cal{A}})+D_A(\rho),
\ee 
where $D_A(\rho)=I_{\cal{A:B}}(\rho)-I_{\cal{A:B}}(\Phi_A(\rho))$ stands for the nonminimized version of the one-way quantum discord (see Ref.~\cite{bilobran15} for further details). In this formulation, it is noticeable that the irreality of $A$ is the sum of the local irreality (that is, the irreality of $A$ given the reduced state $\rho_{\cal{A}}$) with quantum correlations associated with measurements of $A$.

\subsection{Information}
\label{information}

The von Neumann entropy $S(\rho)$ is the quantum-mechanical object widely used to deal with the amount of information associated with the quantum state $\rho$. Here, however, we follow the approach of Ref.~\cite{costa14} and define the amount of information associated with a generic quantum state $\rho$ in a Hilbert space $\cal{H}$ of dimension $d$ as
\be \label{Info}
I(\rho):=\ln{d}-S(\rho).
\ee 
We interpret this as the amount of information available in the reference frame where $\rho$ has been prepared. Clearly, $I$ is maximum (minimum) for a pure (totally mixed) state. In the present approach, therefore, $S(\rho)$ quantifies the ignorance about the state $\rho$.

Consider now a bipartition such that $\cal{H=H_A\otimes H_B}$ and $d=d_{\cal{A}}d_{\cal{B}}=\dim\cal{H}$. It is straightforward to show that
\be \label{I}
I(\rho)=I(\rho_{\cal{A}})+I(\rho_{\cal{B}})+I_{\cal{A:B}}(\rho),
\ee 
where $I(\rho_{\cal{A(B)}})$ is the information related to the subsystem $\cal{A(B)}$, $I_{\cal{A:B}}(\rho)=S(\rho_{\cal{A}})+S(\rho_{\cal{B}})-S(\rho)$ is the mutual information, and $\rho_{\cal{A(B)}}=\text{Tr}_{\cal{B(A)}}\rho$ is the reduced state. The above relation shows that the total information is the sum of local and nonlocal terms, that is, part of the total information is related to the individual subsystems and part is shared by them. The latter term ($I_{\cal{A:B}}$), which is also a measure of the total correlations between $\cal{A}$ and $\cal{B}$, quantifies the information that $\cal{A}$ has about $\cal{B}$, and vice versa. Most importantly, we can check, via unitary invariance of the von Neumann entropy, that in closed systems the total available information is constant, that is $\Delta I=0$. As shown in Ref.~\cite{costa14}, this conservation law allows us to speak of an information flow. For example, when a two-qubit state $\ket{\psi_0}=\ket{a_0}\left(\ket{b_1}+\ket{b_2}\right)/\sqrt{2}$ evolves to $\ket{\psi_t}=\left(\ket{a_1}\ket{b_1}+\ket{a_2}\ket{b_2}\right)/\sqrt{2}$, with $\bra{a_i}a_j\rangle=\bra{b_i}b_j\rangle=\delta_{ij}$, the total information $I=2\ln{2}$, which initially manifested exclusively as local information, is fully transformed into shared information (in this case, entanglement).  

Taking $S_{\cal{A|B}}(\rho)=S(\rho)-S(\rho_{\cal{B}})$ as the definition for the conditional quantum entropy (the entropy of $\cal{A}$ given information about $\cal{B}$) and introducing $I_{\cal{A|B}}(\rho)=\ln{d_{\cal{A}}}-S_{\cal{A|B}}(\rho)$ as the conditional information, we can rewrite Eq. \eqref{I} in the form
\be \label{Icond}
I(\rho)=I(\rho_{\cal{B}})+I_{\cal{A|B}}(\rho),
\ee 
which is particularly interesting for instances where only the part $\cal{B}$ can be accessed. 

\subsection{Strong and weak measurements}
\label{measurement}

One of the basic postulates of quantum mechanics is the state reduction (collapse). It clearly is an effective theoretical tool, a prescription for obtaining the state resulting from a measurement without in any way accounting for the details of the physical interaction with the measurement apparatus. As such, there is no reason {\em a priori} to view the collapse as a real physical phenomenon emerging from the dynamics between the system and the apparatus. In this section we employ this formal perspective. Consider a preparation $\rho\in\cal{H_A\otimes H_B}$ ($\dim\cal{H_{A,B}}=d_{\cal{A,B}}$). According to the quantum axioms, if an operator $A=\sum_aaA_a$, with projectors $A_a=\ket{a}\bra{a}$, is measured in a given run of the experiment and a result $a$ is obtained, then the resulting state is given by
\be 
\cal{C}_{a|A}(\rho):=\frac{(A_a\otimes\mathbbm{1}_{\cal{B}})\,\rho\,(A_a\otimes\mathbbm{1}_{\cal{B}})}{\text{Tr}\left[(A_a\otimes\mathbbm{1}_{\cal{B}})\,\rho\,(A_a\otimes\mathbbm{1}_{\cal{B}})\right]}=A_a\otimes\rho_{\cal{B}|a}.
\ee 
Here $\cal{C}_{a|A}$ is a linear map that formally describes the collapse of the state vector. After a projective measurement of this type, the observer is granted with full information about the reduced state ($\rho_{\cal{A}}=A_a$) of the system. In fact, after the measurement the information about the subsystem $\cal{A}$ reaches its maximum value $I_{\cal{A}}=\ln{d_{\cal{A}}}$. Notice that $\cal{C}_{a|A}^n(\rho)=\cal{C}_{a|A}(\rho)$ for $n\geqslant 1\in\mathbb{Z}$, which correctly implements the condition of repeatability of projective measurements. In addition, we have that $\cal{C}_{a'|A}\cal{C}_{a|A}(\rho)=0$ and $\cal{C}_{a'|A'}\cal{C}_{a|A}(\rho)=\cal{C}_{a'|A'}(\rho)$ for generic (eventually incompatible) observables $A$ and $A'$ acting on $\cal{H_A}$.

We now devise a map that allows us to effectively interpolate between weak and projective measurements. We assume that under the probing process the state $\rho$ is led to
\be \label{Ce}
\cal{C}_{a|A}^{\epsilon}(\rho):=(1-\epsilon)\,\rho+\epsilon\,\cal{C}_{a|A}(\rho),
\ee 
with $\epsilon\in(0,1)$. It is clear that $\cal{C}_{a|A}^{\epsilon}$ represents a strong projective measurement for $\epsilon\to 1$ and no measurement at all for $\epsilon\to~0$. For small $\epsilon$ the map implies just a slightly change in the preparation $\rho$, thus suitably simulating the notion of a weak measurement. Several properties can be derived for the map \eqref{Ce}. First, for $\{A,A'\}$ acting on $\cal{H_A}$ and $B$ acting on $\cal{H_B}$ one has that $[\cal{C}_{a|A}^{\epsilon}(\rho),\cal{C}_{a'|A'}^{\delta}(\rho)]\neq 0$ and $[\cal{C}_{a|A}^{\epsilon}(\rho),\cal{C}_{b|B}^{\delta}(\rho)]=0$. Second, for successive measurements it holds the composition property
\be 
\cal{C}_{a|A}^{\epsilon}\cal{C}_{a|A}^{\delta}=\cal{C}_{a|A}^{\epsilon+\delta-\epsilon\delta}.
\ee  
This allows one to show that $[\cal{C}_{a|A}^{\epsilon}]^n=(1-\epsilon)[\cal{C}_{a|A}^{\epsilon}]^{n-1}+\epsilon\,\cal{C}_{a|A}$ for $n\geqslant 1\in\mathbb{Z}$. Then, via recursion one can prove that
\be 
[\cal{C}_{a|A}^{\epsilon}]^n(\rho)=(1-\epsilon)^n\rho+[1-(1-\epsilon)^n]\,\cal{C}_{a|A}(\rho)=\cal{C}_{a|A}^{1-(1-\epsilon)^n}(\rho),\nonumber
\ee 
which shows that $n$ successive measurements of intensity $\epsilon$ equal a single measurement of intensity $1-(1-\epsilon)^n$. Third, from the above relation we obtain
\be 
\lim_{n\to \infty}[\cal{C}_{a|A}^{\epsilon}]^n=\cal{C}_{a|A},
\ee 
meaning that the action of infinitely many weak measurements is equivalent to a projective measurement. Finally, the relation
\be 
\cal{C}_{a|A}^{\epsilon}(\rho)-\cal{C}_{a|A}^{\delta}(\rho)=(\epsilon-\delta)\left[C_{a|A}(\rho)-\rho \right]
\ee 
provides information about the distance imposed by the application of two measurements of distinct intensities with the same outcome $a$.

\subsection{Monitoring}
\label{monitoring}

In Sec.~\ref{irreality} we used the map $\Phi_A$ as a model for an unrevealed projective measurement. Now we introduce a model that has the capability of interpolating between weak and projective unrevealed measurements. Let us consider a system $\cal{S}$ with a preparation $\rho\in\cal{H_S=H_A\otimes H_B}$. In terms of the eigenbasis $\{a,A_a\}$ of a generic observable $A=\sum_aaA_a$ acting on $\cal{H_A}$, with $A_aA_{a'}=\delta_{aa'}A_a$ and $A_a=\ket{a}\bra{a}$, we can write
\be 
\rho=\sum_{a,a'}\bra{a'}\rho\ket{a}\otimes\ket{a'}\bra{a}=\sum_{a,a'}p_{aa'}\ket{a'}\bra{a}\otimes\rho_{\cal{B}|aa'}.
\ee
Now consider a von Neumann pre-measurement induced by the coupling $H(t)=\epsilon\,g(t)\,A\otimes\mathbbm{1}_{\cal{B}}\otimes P_{\cal{X}}$, where $\cal{X}$ stands for an extra degree of freedom (an ancilla) that will encode the information about $A$, $P_{\cal{X}}$ is the momentum operator acting on $\cal{H_X}$, and $\int_0^t g(t')dt'=1$. By the application of the time-evolution operator $U(t)=\exp{\left[-\tfrac{i}{\hbar}\int_0^t\mathrm{d}t'H(t')\right]}$ on the initial state $\rho\otimes\ket{x_0}\bra{x_0}$ we get the following joint state in $\cal{H_S\otimes H_X}$:
\be 
\rho_{\cal{SX}}(t)=\sum_{a,a'}p_{aa'}\ket{a'}\bra{a}\otimes \ket{x_0+\epsilon\,a'}\bra{x_0+\epsilon\,a}\otimes\rho_{\cal{B}|aa'}.
\ee 
Tracing the ancilla gives
\be
\rho_{\cal{S}}(t)=\textrm{Tr}_{\cal{X}}[\rho_{\cal{SX}}(t)]=\sum_{a,a'}\gamma_{aa'}(\epsilon)\, \ket{a'}\bra{a}\otimes\rho_{\cal{B}|aa'},
\ee
where $\gamma_{aa'}(\epsilon)=\langle x_0+\epsilon\,a|x_0+\epsilon\,a'\rangle$. This term may or may not be small; it depends on the magnitude of the ratio between the distance $\epsilon(a-a')$ and the width of the wave function associated to $\ket{x_0}$.  We then consider the model $\gamma_{aa'}(\epsilon)=(1-\epsilon)+\epsilon\,\delta_{aa'}$ [for $\epsilon\in(0,1)$], which continuously connects a scenario of no interaction $(\epsilon\to 0)$ with a maximally entangling one $(\epsilon\to 1)$. With that, we obtain $\rho_{\cal{S}}(t)=(1-\epsilon)\rho+\epsilon\,\Phi_A(\rho)$. This result leads us to introduce the linear map 
\be \label{Me}
\cal{M}_A^{\epsilon}(\rho):=(1-\epsilon)\,\rho+\epsilon\,\Phi_A(\rho),
\ee 
with $\epsilon\in (0,1)$ and $\Phi_A$ given by Eq.~\eqref{PhiA}. We refer to $\cal{M}_A^{\epsilon}$ as a monitoring with intensity $\epsilon$ of $A$ by $\cal{X}$. [Actually, the relation $\text{Tr}_{\cal{X}}\rho_{\cal{SX}}(t)=\cal{M}_A^{\epsilon}(\rho)$ is a mere expression of Stinespring's dilation theorem \cite{nielsen}.] Notice that $\cal{M}_A^{\epsilon\to 1}(\rho)=\Phi_A(\rho)$. Also, we can write $\cal{M}_A^{\epsilon}(\rho)=\rho-\epsilon[\rho-\Phi_A(\rho)]$, which clearly expresses a degradation of the off-diagonal terms of $\rho$. This is expected since $\cal{M}_A^{\epsilon}$ represents a quantum-noise channel. To see this one can set $K_0=\text{\small $\sqrt{1-\epsilon}$}\,\mathbbm{1}$ and $K_a=\text{\small $\sqrt{\epsilon}$}\,A_a$ and then write $\cal{M}_A^{\epsilon}(\rho)=\sum_{a}K_a\rho K_a^{\dag}$ with $\sum_aK_a^{\dag}K_a+K_0^{\dag}K_0=\mathbbm{1}$, which reveal the operator-sum representation typical of quantum operations~\cite{nielsen}. As such, it is clear that $\cal{M}_A^{\epsilon}$ is a completely positive trace-preserving (CPTP) map. It is also easy to check that $\cal{M}_A^{\epsilon}\cal{M}_B^{\delta}=\cal{M}_B^{\delta}\cal{M}_A^{\epsilon}$ for arbitrary observables $A$ and $B$ and therefore
\be \label{MPhi}
\cal{M}_A^{\epsilon}\Phi_A=\Phi_A\cal{M}_A^{\epsilon}=\Phi_A.
\ee

Now we are in position to formally link measurement with monitoring. When an observer knows that a measurement of $A$ of generic intensity $\epsilon$ has been performed on a preparation $\rho$ but is not informed about the outcome $a$ in a given run of the experiment, the only prediction that can be made by this observer is that the state reduced to $\cal{C}_{a|A}^{\epsilon}(\rho)$ with probability $p_a=\text{Tr}\left[(A_a\otimes\mathbbm{1}_{\cal{B}})\,\rho\right]$. Without information about the specific outcome $a$, it follows from the definition \eqref{Ce} that the better prediction the observer can make is 
\be 
\sum_ap_a\cal{C}_{a|A}^{\epsilon}(\rho)=\cal{M}_A^{\epsilon}(\rho).
\ee 
Conceptually, there is an important point to make, namely, that monitoring is indistinguishable from an unrevealed measurement. The left-hand side of the above relation was constructed with the basis on a measurement [eventually a collapsing one (for $\epsilon\to 1$)] that has been secretly conducted. The right-hand side, in its turn, was derived via an entangling dynamics with an ancilla, without any {\em a priori} link with the state reduction. This points out that unrevealed collapse is formally equivalent to entanglement plus discard, which suggests that the state vector reduction can be interpreted as information updating rather than as a physical reduction of the state vector. We will return to this point later within an informational perspective.

We now derive the mathematical properties of $\cal{M}_A^{\epsilon}$. Concerning successive applications of the map, one shows from \eqref{Me} and \eqref{MPhi} that $[\cal{M}_A^{\epsilon}]^n(\rho)=(1-\epsilon)[\cal{M}_A^{\epsilon}]^{n-1}(\rho)+\epsilon\Phi_A(\rho)$, for $n\geqslant 1\in\mathbb{Z}$. By recursion one obtains that
\be \label{Men}
[\cal{M}_A^{\epsilon}]^n(\rho)=(1-\epsilon)^n\rho+\big[1-(1-\epsilon)^n\big]\Phi_A(\rho).
\ee
Notice that $[\cal{M}_A^{\epsilon\to 1}]^n(\rho)=\Phi_A(\rho)$, as expected. Also, one has that $[\cal{M}_A^{\epsilon}]^n(\rho)=\cal{M}_A^{1-(1-\epsilon)^n}(\rho)$, which shows that $n$ monitorings of intensity $\epsilon$ is equivalent to a single monitoring of intensity $1-(1-\epsilon)^n$. Finally,
\be
\lim\limits_{n\to\infty}[\cal{M}_A^{\epsilon}]^n=\Phi_A,
\ee
which means that infinitely many weak monitorings, executed either sequentially or simultaneously, establish the reality of the monitored observable for any state. From the above  relations, further composition properties can be derived:
\begin{subequations}
\beq 
&&\lim_{n\to\infty}[\cal{M}_A^{\epsilon/n}]^n=\cal{M}_A^{1-e^{-\epsilon}},\\
&&\cal{M}_A^{\delta}\cal{M}_A^{\epsilon}=\cal{M}_A^{\delta+\epsilon-\delta\epsilon}.
\eeq 
\end{subequations}
for $\{\epsilon,\delta\}\in(0,1]$ and $n\geqslant 1\in\mathbb{Z}$. Also, by noticing that $\cal{M}_A^{\epsilon}(\rho)-\rho=\epsilon[\Phi_A(\rho)-\rho]$, one shows that
\be\label{M-M}
\cal{M}_A^{\epsilon}(\rho)-\cal{M}_A^{\delta}(\rho)=(\epsilon-\delta)[\Phi_A(\rho)-\rho].
\ee 
%

\section{Measurement, information and reality}
\label{results}

We are now ready to present the main contribution of this paper, namely, the formal development of connections between the notion of reality, measurement, and information. In our view, such a task has not been accomplished so far due to the lack of a formal quantifier of reality, which is now available (see Sec.~\ref{irreality} and Ref.~\cite{bilobran15}).

\subsection{Monitoring increases reality}
\label{mir}

Consider a preparation $\rho\in\cal{H_A\otimes H_B}$. For this state, the degree of irreality of a generic observable $A$ is given by $\mathfrak{I}(A|\rho)$. Under a monitoring $\cal{M}_A^{\epsilon}$ of arbitrary intensity $\epsilon$, the irreality of $A$ changes to $\mathfrak{I}(A|\cal{M}_A^{\epsilon}(\rho))$. Although a quantifier $\mathfrak{R}$ of reality itself has not been defined, it is clear that this concept should be dual to irreality, that is, $\Delta\mathfrak{I}(A)+\Delta\mathfrak{R}(A)=0$. Then, under the monitoring $\cal{M}_A^{\epsilon}$ on $\rho$, the reality of $A$ changes as
\be \label{DRI}
\Delta\mathfrak{R}(A):=-\Delta \mathfrak{I}(A)=\mathfrak{I}(A|\rho)-\mathfrak{I}(A|\cal{M}_A^{\epsilon}(\rho)).
\ee 
Using the definition of irreality \eqref{frakI} and the hierarchy \eqref{MPhi}, it follows that
\be \label{DR}
\Delta \mathfrak{R}(A)=S(\cal{M}_A^{\epsilon}(\rho))-S(\rho),
\ee 
which is a non-negative quantity. If $\rho=\Phi_A(\rho)$, then $\Delta\mathfrak{R}=0$, since in this case the preparation $\rho$ is already a state of reality for $A$. If $\epsilon\to 1$, then the reality change saturates to its maximum value $\Delta\mathfrak{R}_{\max}(A)=\mathfrak{I}(A|\rho)$, meaning that the reality increases precisely by the value that defined the amount by which the observable was unreal. From the concavity of the von Neumann entropy and the non-negativity of irreality we obtain
\be\label{DRA>0}
\Delta \mathfrak{R}(A)\geqslant \epsilon\,\mathfrak{I}(A|\rho),
\ee
with the equality holding for $\epsilon\to 1$. [Actually, the equality also holds for $\epsilon\to 0$ and $\rho=\Phi_A(\rho)$, but in these cases both the left-hand-side and right-hand-side terms vanish.] Hence, apart from extremal instances, the reality of an observable increases under monitoring. Furthermore, one shows that under monitoring the reality increase is bounded from above. Take Fannes's inequality $|S(\rho)-S(\sigma)|\leqslant T\ln{(d-1)}+H(T)$~\cite{audenaert07}, where $T(\rho,\sigma)=\tfrac{1}{2}\text{Tr}||\rho-\sigma||_1 \in [0,1]$ is the trace norm, $H(T)=-T\ln{T}-(1-T)\ln{(1-T)}$ is the Shannon entropy, $||\varrho||_1=(\varrho^{\dag}\varrho)^{1/2}$ is the Schatten 1-norm, and $d=\dim{\cal{H}}$. Using the relation \eqref{M-M}, one shows that $T(\cal{M}_A^{\epsilon}(\rho),\rho)=\epsilon\,\tau$, with $\tau\equiv T(\Phi_A(\rho),\rho)$. We then arrive at
\be \label{DR<ub}
\Delta\mathfrak{R}(A)\leqslant \epsilon\,\tau\ln{(d-1)}+H\left(\epsilon\,\tau\right).
\ee 
It can be checked for $\tau>0$ that the above upper bound can never reach the value $d\sqrt{\epsilon\,\tau/e}$, which can therefore be taken as a simpler estimate for the $\Delta\mathfrak{R}(A)$ upper bound. The inequalities \eqref{DRA>0} and \eqref{DR<ub} define our first result: A monitoring of intensity $\epsilon$, which can be interpreted either as an unrevealed measurement or as an operation involving entanglement plus discard, implies a finite increase not less than $\epsilon\mathfrak{I}(A|\rho)$ in the reality of the monitored observable. Notice that the increment in the reality, whose upper bound is regulated by the monitoring intensity $\epsilon$, can be made to be infinitesimal.

\subsection{Monitoring increases the reality of incompatible observables}
\label{mirio}

After measuring $\sigma_z$ for a spin-$\tfrac{1}{2}$ particle prepared in a generic state $\rho$ and announcing the result, the state of the system collapses to one of the states $\ket{\pm z}=\left(\ket{+x}\pm\ket{-x} \right)/\sqrt{2}$. Thus, while the reality of $\sigma_z$ increases in the process, the irreality of an incompatible observable, say, $\sigma_x$, reaches its maximum value, so its reality decreases. As we show now, the situation is rather different as a monitoring is involved.

Let $A$ and $A'$ be incompatible observables acting on $\cal{H_A}$. We want to see how the reality of $A'$ changes when a monitoring $\cal{M}_A^{\epsilon}$ of $A$ is performed on $\rho\in\cal{H_S}$. Via the relations~\eqref{frakI} and \eqref{DRI}, the reality change $\Delta\mathfrak{R}(A')=\mathfrak{I}(A'|\rho)-\mathfrak{I}(A'|\cal{M}_A^{\epsilon}(\rho))$ can be written in the form
\be \label{DRA'}
\Delta\mathfrak{R}(A')=S(\Phi_{A'}(\rho))+S(\cal{M}_A^{\epsilon}(\rho))-S(\rho)-S(\Phi_{A'}\cal{M}_A^{\epsilon}(\rho)).
\ee 
To infer the behavior of this quantity, we consider an extended space $\cal{H_S\otimes H_X\otimes H_Y}$, with $\cal{H_S=H_A\otimes H_B}$, and write
\be \label{rhoSXY}
\rho_{\cal{SXY}}=U_{\cal{SX}}U_{\cal{SY}}\,\Big(\rho\otimes\ket{x_0}\bra{x_0}\otimes\ket{y_0}\bra{y_0}\Big)\,U_{\cal{SY}}^{\dag}U_{\cal{SX}}^{\dag},
\ee 
with unitary transformations such that
\be 
U_{\cal{SX}}=e^{-\frac{i\epsilon}{\hbar}A\otimes\mathbbm{1}_{\cal{B}}\otimes P_{\cal{X}}},\qquad 
U_{\cal{SY}}=e^{-\frac{i\delta}{\hbar}A'\otimes\mathbbm{1}_{\cal{B}}\otimes P_{\cal{Y}}},
\ee
and $[U_{\cal{SX}},U_{\cal{SY}}]\neq 0$. These operators refer to von Neumann pre measurements of the observables $A$ and $A'$, with intensities $\epsilon$ and $\delta$, via ancillary systems $\cal{X}$ and $\cal{Y}$, respectively. From the relations above and the Stinespring dilation theorem [see also Eq.~\eqref{Me}] one may directly obtain the reduced state $\rho_{\cal{SX}}=U_{\cal{SX}}\left(\cal{M}_{A'}^{\delta}(\rho)\otimes\ket{x_0}\bra{x_0}\right)U_{\cal{SX}}^{\dag}$, which by unitary invariance of the von Neumann entropy implies that $S(\rho_{\cal{SX}})=S(\cal{M}_{A'}^{\delta}(\rho))$. For the same reason, $S(\rho_{\cal{SXY}})=S(\rho)$. To compute the reduction $\rho_{\cal{SY}}$ we first note that
\beq
U_{\cal{SX}}U_{\cal{SY}}=U_{\cal{SX}}\left(e^{-\frac{i\delta}{\hbar}A'\otimes\mathbbm{1}_{\cal{B}}\otimes P_{\cal{Y}}}\right)U_{\cal{SX}}^{\dag}U_{\cal{SX}}=e^{-\frac{i\delta}{\hbar}\tilde{A}'\otimes\mathbbm{1}_{\cal{B}}\otimes P_{\cal{Y}}}U_{\cal{SX}},\nonumber 
\eeq
where $\tilde{A}'\otimes\mathbbm{1}_{\cal{B}}=U_{\cal{SX}}(A'\otimes\mathbbm{1}_{\cal{B}})U_{\cal{SX}}^{\dag}$. Because $\tilde{A}'$ is Hermitian, we have thus shown that $U_{\cal{SX}}U_{\cal{SY}}=\tilde{U}_{\cal{SY}}U_{\cal{SX}}$, with a new unitary operator $\tilde{U}_{\cal{SY}}$. With this result, we can turn to Eq.~ \eqref{rhoSXY} to show that $\rho_{\cal{SY}}=\tilde{U}_{\cal{SY}}\left(\cal{M}_A^{\epsilon}(\rho)\otimes\ket{y_0}\bra{y_0}\right)\tilde{U}_{\cal{SY}}^{\dag}$. It thus follows that $S(\rho_{\cal{SY}})=S(\cal{M}_A^{\epsilon}(\rho))$. From all this, it also emerges that $\rho_{\cal{S}}=\cal{M}_A^{\epsilon}\cal{M}_{A'}^{\delta}(\rho)=\cal{M}_{A'}^{\delta}\cal{M}_A^{\epsilon}(\rho)$. Then, from the strong subadditivity of the von Neumann entropy [$S(\rho_{\cal{SXY}})+S(\rho_{\cal{S}})\leqslant S(\rho_{\cal{SX}})+S(\rho_{\cal{SY}})$] we arrive at
\be 
S(\rho)+S(\cal{M}_{A'}^{\delta}\cal{M}_A^{\epsilon}(\rho))\leqslant S(\cal{M}_{A'}^{\delta}(\rho))+S(\cal{M}_{A}^{\epsilon}(\rho)).
\ee 
Given that $\cal{M}_{A'}^{\delta\to 1}=\Phi_{A'}$, we return to Eq. \eqref{DRA'} to obtain
\be \label{DRA'>0}
\Delta\mathfrak{R}(A')\geqslant 0,
\ee 
with equality holding for $\{\epsilon,\delta\}\to 0,1$ and $\rho=\Phi_{A(A')}(\rho)$. This result is surprising, as it shows that under monitoring of $A$ the reality of $A'$ will also increase in general. In fact, along with the inequality \eqref{DRA>0}, this shows that a monitoring typically increases the global reality of a system.

It is worth mentioning that the inequalities \eqref{DRA>0} and \eqref{DRA'>0}, along with some results reported in Ref. \cite{bilobran15}, prove the monotonicity of BA's irreality under monitoring (a CPTP map), that is, $\mathfrak{I}(A|\rho)\geqslant \mathfrak{I}(A|\cal{M}_O^{\epsilon}(\rho))$ for $\rho\in\cal{H_A\otimes H_B}$ and $O$ being a generic Hermitian operator acting on $\cal{H_A}$ or $\cal{H_B}$ and $A$ an Hermitian operator acting on $\cal{H_A}$. This means that the irreality never increases under monitoring. This observation naturally raises the following question: Is there any scenario in which the irreality of an observable can increase? Next we address this question.

\subsection{Generation of irreality}
\label{goi}

Consider two maximally incompatible observables $A$ and $A'$ acting on $\cal{H_A}$, meaning that their eigenstates constitute mutually unbiased bases satisfying $|\langle a|a'\rangle|^2=1/d_{\cal{A}}$, where $d_{\cal{A}}=\dim\cal{H_A}$. Let $\rho_{[A']}$ denote a reality state for $A'$, that is, $\rho_{[A']}=\Phi_{A'}(\rho)$ and therefore $\mathfrak{I}(A'|\rho_{[A']})=0$. Under monitoring of the incompatible observable $A$ the state transforms to $\cal{M}_{A}^{\epsilon}(\rho_{[A']})=(1-\epsilon)\Phi_{A'}(\rho)+\tfrac{\mathbbm{1}}{d_{\cal{A}}}\otimes\rho_{\cal{B}}$, where $\rho_{\cal{B}}=\text{Tr}_{\cal{A}}(\rho)$. Since this state does not change under $\Phi_{A'}$ we can check that $\mathfrak{I}(A'|\cal{M}_{A}^{\epsilon}(\rho_{[A']}))=0$. This result shows that the monitoring of $A$ does not increase the irreality of the incompatible observable $A'$. As such, it is an illustration of the more general result obtained in the preceding section. Interestingly, now we show that this situation changes when revealed measurements are involved. To this end, let us invoke the map $\cal{C}_{a|A}^{\epsilon}$, which was introduced in Eq.~\eqref{Ce} as an effective descriptor for a measurement of $A$ with generic intensity $\epsilon$ and known outcome $a$. In this case, we have $\cal{C}_{a|A}^{\epsilon}(\rho_{[A']})=(1-\epsilon)\Phi_{A'}(\rho)+\epsilon A_a\otimes\rho_{\cal{B}}$, which {\em does} change under the map $\Phi_{A'}$ since $\Phi_{A'}(A_a)=\tfrac{\mathbbm{1}}{d_{\cal{A}}}$. It then follows that $\mathfrak{I}(A'|\cal{C}_{a|A}^{\epsilon}\rho_{[A']})>0$, that is, the irreality of $A'$ indeed increases under revealed measurements. In addition, by direct application of Fannes's inequality [see the inequality \eqref{DR<ub} and its derivation] one may show that
\be 
\mathfrak{I}(A'|\cal{C}_{a|A}^{\epsilon}\rho_{[A']})\leqslant \epsilon\,\tilde{\tau} \ln{(d-1)+H(\epsilon\,\tilde{\tau})},
\ee 
where $\tilde{\tau}\equiv T\left(\tfrac{\mathbbm{1}}{d_{\cal{A}}},A_a\right)=1-1/d_{\cal{A}}$. Again we can take $d\sqrt{\epsilon\,\tilde{\tau}/e}$ as a simpler estimate for the upper bound given above. We have thus proved that irreality can be generated for $A'$ by means of revealed measurements of the incompatible observable $A$. Being controlled by the measurement intensity $\epsilon$, it is clear that the generated irreality can be made arbitrarily small.

We now assess the possibility of generating irreality in unitary dynamics. Consider a preparation $\rho\in\cal{H_A\otimes H_B}$. Let $U_{\cal{B}}$ be a unitary transformation acting on $\cal{H_B}$. Since $\Phi_A$ commutes with $U_{\cal{B}}$ it follows that $\mathfrak{I}(A|U_{\cal{B}}\rho U_{\cal{B}}^{\dag})=\mathfrak{I}(A|\rho)$. This shows that a local unitary transformation is not able to promote an increase of irreality in a remote site. We are left then with global unitary transformations. In what follows we will conduct our analysis in terms of a concrete example involving the frontal scattering of a particle of mass $m$, initially prepared in a Gaussian wave packet of mean momentum $p_0=mv_0$ and width $\Delta p=m\Delta v$, by a molecule of mass $M$, prepared in a Gaussian wave packet of null mean momentum and width $\Delta P=M\Delta v$. Assuming that the probability of the scattering to occur is 1/2 and that the collision is elastic, then the nonrelativistic energy and momentum conservation laws along with a unitary evolution require, up to a normalization factor, that
\be \label{scatt}
\ket{p_0}\ket{0}\to\ket{p_0}\ket{0}+\ket{(1-\alpha) p_0}\ket{\alpha p_0},
\ee 
where $\alpha=2/(1+\xi)$ and $\xi=m/M$. The notation is such that ``$\ket{p}\ket{P}$'' represents a product of wave packets with mean momentum $p$ and variance $(\Delta p)^2$ for the particle and $P$ and $(\Delta P)^2$ for the molecule, respectively. Via direct calculations one computes the overlaps:
\begin{subequations}
\beq 
O_{\text{part}}&\equiv&\big|\bra{p_0}(1-\alpha)p_0\rangle\big|=\exp{\left[-\tfrac{1}{2}\left(\frac{1}{1+\xi}\frac{v_0}{\Delta v}\right)^2\right]}, \\
O_{\text{mol}}&\equiv&\big|\bra{0}\alpha p_0\rangle\big|=\exp{\left[-\tfrac{1}{2}\left(\frac{\xi}{1+\xi}\frac{v_0}{\Delta v}\right)^2\right]}.
\eeq 
\end{subequations}

Now, since a measure of irreality for continuous variables is not yet available in the literature, here we approximately treat position and momentum as discrete variables relative to some (experimental) resolutions $\delta_x$ and $\delta_p$ and then apply the present formalism. Within this framework, if $\Delta p<\delta_p$, then the initial momentum of the particle is effectively real. Let us also assume that $\Delta v\ll v_0$ and consider two regimes. First, if the molecule is not so heavy, so that $\xi\approx 1$, then $O_{\text{part}}\approx O_{\text{mol}}\approx 0$ and the state \eqref{scatt} is highly entangled. The relation \eqref{LIQC} implies, as a consequence of the quantum correlations generated by the scattering, that the irreality of the momentum of the particle has increased. This shows that an entangling unitary dynamics is an effective mechanism to create irreality. On the other hand, if we restrict ourselves to the subsystem particle and thus trace out the molecule degree of freedom, then the resulting reduced state will be the mixture $\ket{p_0}\bra{p_0}+\ket{0}\bra{0}$, which means no irreality whatsoever. Hence, as far as the particle is considered as an individual, there is no increase in the irreality of its momentum. We then move to the second regime of interest. Consider now a very heavy molecule, so that $\xi\to0$. In this case, $O_{\text{part}}\approx 0$, $O_{\text{mol}}\to 1$, and therefore $\ket{0}\approx\ket{2\alpha p_0}$, meaning that the state of system evolves from $\ket{p_0}\ket{0}$ to $(\ket{p_0}+\ket{-p_0})\ket{0}$. In other words, while no entanglement is produced between the subsystems, a significant quantum superposition is created. In this case, the local irreality noticeably increases. Notice that because $\ket{0}\approx\ket{2\alpha p_0}$ the time evolution of the global state is such that the momentum conservation seems to have been effectively frustrated. 

This mechanism also appears in paradigmatic experiments where local irreality (coherence) is generated. When a particle initially moving with a well-defined momentum $p\,\hat{\mathrm{x}}$ diffracts through an orifice (a tiny circular slit) it ends up in a superposition of momentum states associated with directions orthogonal to $\hat{\mathrm{x}}$. In this case, since we cannot detect any motion of the orifice, which is rigidly attached to the laboratory (the reference frame), we have an effective frustration of the momentum conservation law. The situation is similar when the spin of a particle is flipped by a magnet which, being fixed in the laboratory, cannot rotate relatively to this reference frame. Then the observer perceives an effective violation of the total angular momentum conservation. These examples suggest that the frustration of a conservation law within a unitary dynamics is {\em the} crucial mechanism for the generation of local irreality in interacting dynamics.

\subsection{Information-reality duality}
\label{ird}

\begin{figure}[htb]
\centerline{\includegraphics[scale=0.09]{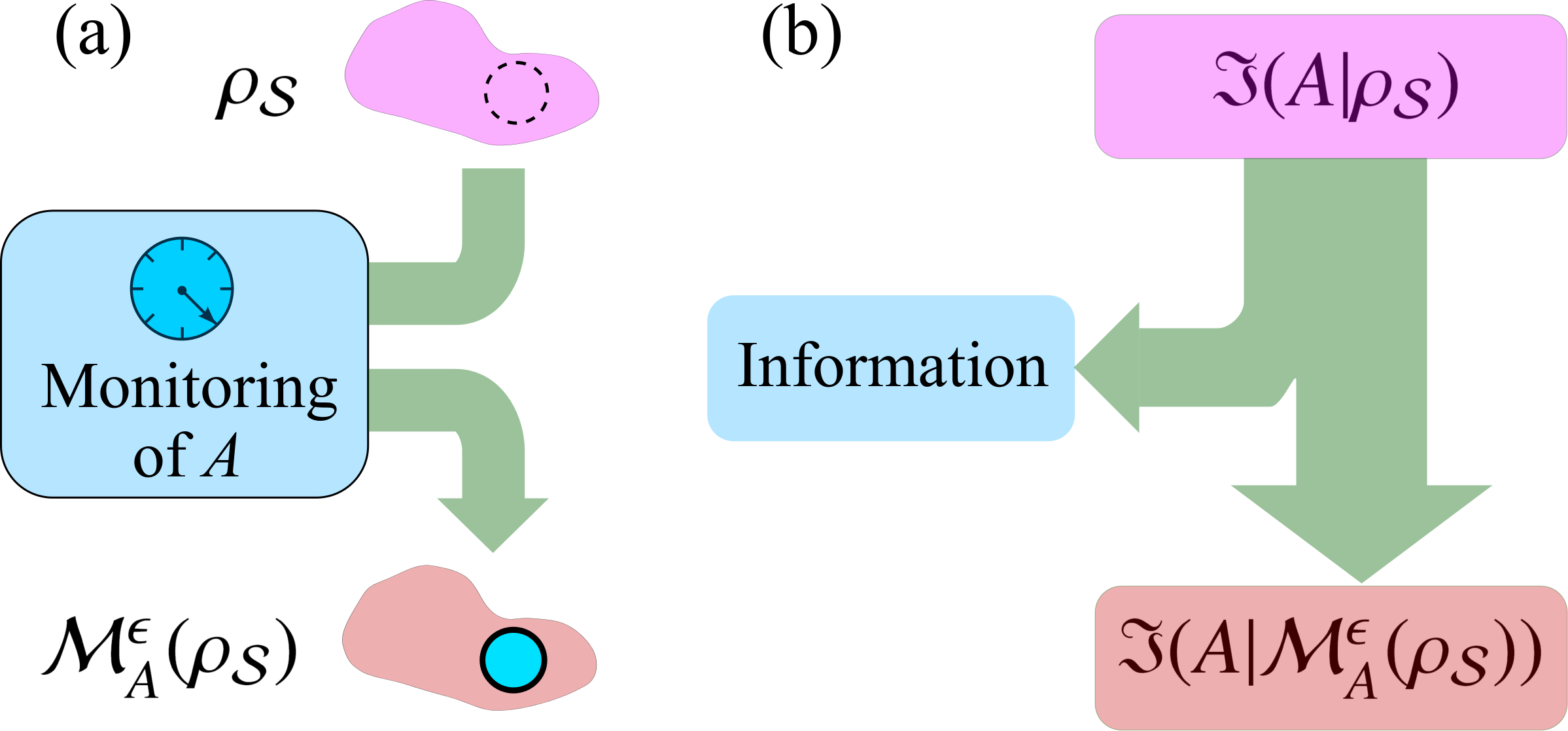}}
\caption{(Color online) (a) Generic state $\rho_{\cal{S}}\in\cal{H_A\otimes H_B}$ becomes $\cal{M}_A^{\epsilon}(\rho_{\cal{S}})$ under monitoring of an observable $A$ acting on $\cal{H_A}$. (b) Same process abstractly pictured in terms of irreality and information. As both local and global information is generated, the irreality of $A$ decreases [see Eq.~\eqref{complI}]. }
\label{fig1}
\end{figure}

A particularly interesting aspect that emerges in the present framework is a clear link between information and reality. Consider an instance in which a system $\cal{S}$ initially prepared in a state $\rho_{\cal{S}}\in\cal{H_A\otimes H_B}$ ends up in $\cal{M}_A^{\epsilon}(\rho_{\cal{S}})$ after the monitoring of a generic observable $A$ acting on $\cal{H_A}$. As mentioned above, the Stinespring theorem ensures that this mapping can be cast in terms of an entangling dynamics $U(t)$ between $\cal{S}$ and some extra degree of freedom $\cal{X}$ initially prepared in a state $\ket{x_0}\bra{x_0}$, that is,
\be \label{rhoSt}
\cal{M}_A^{\epsilon}(\rho_{\cal{S}})=\text{Tr}_{\cal{X}}\left[U(t)\,\rho_{\cal{S}}\otimes\ket{x_0}\bra{x_0}\,U^{\dag}(t) \right]=\rho_{\cal{S}}(t).
\ee 
The mutual information of the joint system $\cal{SX}$ at an arbitrary instant $t$ reads $I_{\cal{S:X}}(t)=S(\rho_{\cal{S}}(t))+S(\rho_{\cal{X}}(t))-S(\rho_{\cal{SX}}(t))$. Since the joint evolution is unitary, then $S(\rho_{\cal{SX}}(t))=S(\rho_{\cal{SX}}(0))$. Introducing $\Delta S_{\cal{S(X)}}=S(\rho_{\cal{S(X)}}(t))-S(\rho_{\cal{S(X)}}(0))$, the change of the mutual information with time reads $\Delta I_{\cal{S:X}}=\Delta S_{\cal{S}}+\Delta S_{\cal{X}}$. Via $I_{\cal{S(X)}}=\ln{d_{\cal{S(X)}}}-S(\rho_{\cal{S(X)}})$ and Eq.~\eqref{rhoSt} we respectively have $\Delta S_{\cal{X}}=-\Delta I_{\cal{X}}$ and $\Delta S_{\cal{S}}=S(\cal{M}_A^{\epsilon}(\rho_{\cal{S}}))-S(\rho_{\cal{S}})$, so that $\Delta I_{\cal{S:X}}+\Delta I_{\cal{X}}=S(\cal{M}_A^{\epsilon}(\rho_{\cal{S}}))-S(\rho_{\cal{S}})$. Using Eq.~\eqref{DRI} we then arrive at
\be \label{complI}
\Delta\left(I_{\cal{S:X}}+I_{\cal{X}}\right)+\Delta\mathfrak{I}(A)=0.
\ee 
From the identity~\eqref{I} and the unitarity of the joint dynamics it follows that $\Delta\left(I_{\cal{S:X}}+I_{\cal{X}}\right)=-\Delta I_{\cal{S}}$, which allows us to write
\be \label{complR}
\Delta I_{\cal{S}}+\Delta\mathfrak{R}(A)=0.
\ee 
The relations \eqref{complI} and \eqref{complR} formally state the complementarity between (ir)reality and information, which is another important contribution of this work. As is schematically illustrated in Fig.~\ref{fig1}, variations in both the local information $I_{\cal{X}}$ associated with the subsystem $\cal{X}$ and the information $I_{S:X}$ shared by $\cal{S}$ and $\cal{X}$ directly imply variations in $A$'s irreality. In particular, it is interesting to note that if $\rho_{\cal{S}}$ is a pure state, then the joint initial state is pure as well and the entanglement $E$ in the system $\cal{SX}$ is given by $E=S(\rho_{\cal{S}(X)}(t))$. Since $\Delta I_{\cal{S}}=I_{\cal{S}}(t)-I_{\cal{S}}(0)=-E$, it follows that
\be 
\Delta\mathfrak{R}(A)=E,
\ee 
which explicitly shows that the reality change in $A$ is determined by the amount of entanglement between $\cal{S}$ and $\cal{X}$. In other words, because $\cal{X}$ gets information about $A$, this observable becomes real. This is in full agreement with the results reported in Ref. \cite{angelo15}, where entanglement is shown to prevent the wavelike behavior of a quantum system.

\subsection{The measurement problem}
\label{mp}

The foundational relevance of the measurement problem needs no emphasis. Here we hope to shed some light on this longstanding issue by using the tools introduced above. We consider the well-known drama proposed by Everett \cite{everett57}, in which an external observer describes a measurement conducted within a laboratory by an internal observer. The conflict emerges as we note that for the internal observer an irreversible state reduction occurs, whereas for the external one, who can conceive of the internal observer as part of a physical system, only a reversible dynamics takes place. 

To approach this puzzle we take an informational perspective and consider, from the viewpoint of the external observer $\cal{O}_{\text{ext}}$, three physical systems, namely, the internal observer $\cal{O}$, an apparatus $\cal{A}$, and a system of interest $\cal{S}$. These systems are described quantum mechanically by $\cal{O}_{\text{ext}}$, who naturally does not include oneself in the description. Let $\sigma_{\cal{AS}}$ be the joint state after the apparatus has got correlated with the system. In the last stage of the measurement process, $\cal{O}$ looks at the apparatus, that is, indirectly interacts with $\cal{A}$ by means of the photons scattered by $\cal{A}$. Without interacting with the joint system $\cal{OAS}$, $\cal{O}_{\text{ext}}$ describes the dynamics in terms of the unitary evolution $\rho_{\cal{OAS}}=U_{\cal{OA}}\left(\sigma_{\cal{AS}}\otimes\ket{o}\bra{o}\right)U_{\cal{OA}}^{\dag}$. According to Eq.~\eqref{I}, the information is distributed over the system as 
\be 
I(\rho_{\cal{OAS}})=I(\rho_{\cal{O}})+I_{\cal{O:AS}}(\rho_{\cal{OAS}})+I(\rho_{\cal{AS}}).
\ee 
The first two terms on the right-hand side refer to information that cannot be accessed by $\cal{O}$, as they refer to the state of $\cal{O}$ and its correlation with the part $\cal{AS}$ as seen from the perspective of $\cal{O}_{\text{ext}}$. This is an irremovable limitation because $\cal{O}$ cannot ascribe a quantum state for oneself and therefore has no way to assess the terms $I(\rho_{\cal{O}})+I_{\cal{O:AS}}(\rho_{\cal{OAS}})$. Let us doubly emphasize this point by recalling that no reference frame can describe its own physical state. By its turn, the third term on the right-hand side can be written, according to Eq.~\eqref{Icond}, as $I(\rho_{\cal{AS}})=I(\rho_{\cal{A}})+I_{\cal{S|A}}(\rho_{\cal{AS}})$, where $I_{\cal{S|A}}$ is expected to be the only informational content that $\cal{O}$ can obtain about $\cal{S}$ through the measurement process. To see that this is indeed the case, let us move to $\cal{O}$'s reference frame, wherein the unitary evolution $U_{\cal{OA}}$ is not applicable. According to the reduction postulate, upon collection of scattered photons, $\cal{O}$ will (somehow) perceive a state $\cal{C}_{a|A}(\sigma_{\cal{AS}})=A_a\otimes\sigma_{\cal{S}|a}$ in a particular run of the experiment. The average entropy associated with many runs will be $\bar{S}_{\cal{S|A}}=\sum_ap_aS(\cal{C}_{a|A}(\sigma_{\cal{AS}}))=\sum_ap_aS(\sigma_{\cal{S}|a})$. Using the joint entropy theorem \cite{nielsen}, one shows that this result can be written as $\bar{S}_{\cal{S|A}}=S(\Phi_A(\sigma_{\cal{AS}}))-S(\Phi_A(\sigma_{\cal{A}}))=S_{\cal{S|A}}(\Phi_A(\sigma_{\cal{AS}}))$, which refers to the remaining ignorance about $\cal{S}$ given that $\cal{A}$ has been accessed and collapsed. The average information acquired by $\cal{O}$ about $\cal{S}$ through the observation of $\cal{A}$ is, by definition, $\bar{I}_{\cal{S|A}}:=\ln{d_{\cal{S}}}-\bar{S}_{\cal{S|A}}$. It can be written as
\be 
\bar{I}_{\cal{S|A}}=\ln{d_{\cal{S}}}-\sum_ap_aS(\cal{C}_{a|A}(\sigma_{\cal{AS}})).
\ee 
To compute $I_{\cal{S|A}}$, the information that $\cal{O}$ can access about $\cal{S}$ via interaction with $\cal{A}$, from $\cal{O}_{\text{ext}}$'s perspective, we apply the Stinespring theorem to write $\rho_{\cal{AS}}=\text{Tr}_{\cal{O}}\rho_{\cal{OAS}}=\Phi_A(\sigma_{\cal{AS}})$, which presumes that a strong monitoring has occurred inside $\cal{O}$'s laboratory. It follows from the definition of conditional information that $I_{\cal{S|A}}=\ln{d_{\cal{S}}}-S_{\cal{S|A}}(\Phi_A(\sigma_{\cal{AS}}))=\ln{d_{\cal{S}}}-\bar{S}_{\cal{S|A}}$, which implies that $I_{\cal{S|A}}=\bar{I}_{\cal{S|A}}$, as we wanted to prove. This result can also be written as
\be
I_{\cal{S|A}}(\Phi_A(\sigma_{\cal{AS}}))=\ln{d}_{\cal{S}}-\sum_ap_aS(\cal{C}_{a|A}(\sigma_{\cal{AS}})),
\ee  
which explicitly states the link between the information related to an unread measurement, as signalized by $\Phi_A$, with the information collected through several reductions of the form $\cal{C}_{a|A}$.
The main message here is that the amount of information acquired by $\cal{O}$ about $\cal{S}$ is always the same regardless of the reference frame we choose to assess it. In $\cal{O}$'s frame we use the notion of state collapse and compute an average information, whereas in $\cal{O}_{\text{ext}}$'s frame the same informational content is obtained by considering a unitary evolution plus the discard of $\cal{O}$. From this point of view, therefore, there is no paradox. It is clear, however, that because information flows from $\cal{AS}$ to $\cal{O}$, this observer can in no way, in one's reference frame, deal with an information-preserving dynamics. In other words, in one's perspective the entropy of $\cal{AS}$ always decreases. 

There is another involving aspect of the measurement problem that needs attention, namely, the occurrence of individual outcomes $\cal{C}_{a|A}(\sigma_{\cal{AS}})=A_a\otimes\sigma_{\cal{S}|a}$ from $\cal{O}$'s perspective in each run of the experiment. This is no doubt a major difficulty around the issue. To discuss this point we focus on a concrete example where the $z$ component of spin is measured for a spin-1/2 particle in a preparation $\alpha\ket{+}+\beta\ket{-}$. In the first stage of the experiment, the spin degree of freedom gets correlated, via a Stern-Gerlach field, with the spatial coordinate $z$ of the particle. The resulting state can be written in the form $\ket{\psi_{\cal{S}}}=\alpha\ket{+}\ket{+\bar{z}}+\beta\ket{-}\ket{-\bar{z}}\in\cal{H_S}$, where $\langle z| \pm\bar{z}\rangle \equiv \psi(z\mp \bar{z})$ stands for a probability amplitude centered at $\pm\bar{z}$. For the role of apparatus we imagine a detection array composed of ideally tiny detectors that get visible marks (via some ionizing process) upon absorption of a particle. The $i$-th detector starts in a state $\ket{\phi_i,\varepsilon}=\int\mathrm{d}z\,\phi(z-z_i)\ket{z}\ket{\varepsilon}$, where $|\phi(z-z_i)|^2$ is assumed to be a very sharp normalized Gaussian distribution of width $\delta z$ centered at $z_i$ and $\ket{\varepsilon}$ is a state of energy such that $\varepsilon=e$ (excited) when a mark appears in the detector and $\varepsilon=g$ (ground) otherwise. In our model, $\langle \varepsilon|\varepsilon'\rangle=\delta_{\varepsilon,\varepsilon'}$ and $\langle \phi_i|\phi_j\rangle=\exp\big[-(z_i-z_j)^2/(8\delta z^2)\big]\approx \delta_{z_i,z_j}$, meaning that any two detectors and their signs are distinguishable, which is a desirable feature of any detection system. Given the finite size of the detectors, one can consistently work with a discretized space for the particle, where $\langle z_i|z_j\rangle\approx\delta_{i,j}/\delta z$ so that
\be 
\ket{\pm\bar{z}}=\int\mathrm{d}z\,\psi(z)\,\ket{z\pm\bar{z}}\approx \sum_k\delta z\,\psi(z_k)\,\ket{z_k\pm\bar{z}},
\ee 
with $\bar{z}=n\,\delta z$ for $n\in\mathbb{Z}$. By virtue of the space discretization, one has $z_k=k\,\delta z$ and therefore $z_k\pm\bar{z}=z_{k\pm n}$. Now let $\ket{\psi_{\cal{A}}}=\bigotimes_i\ket{\phi_i,g}$ be the initial state of the apparatus. Our model admits that upon physical interactions one has that $\ket{z_k}\ket{\psi_{\cal{A}}}\to \ket{z_k}\ket{1_k}$, where we have introduced the one-excitation state $\ket{1_k}\equiv\ket{\phi_k,e}\bigotimes_{i\neq k}\ket{\phi_i,g}$ with $\langle 1_k|1_{k'}\rangle\approx\delta_{k,k'}$, which means that the detector at $z_k$ gets excited whereas all the others remain unexcited. By use of this model, the initial joint state
\be 
\ket{\psi_{\cal{S}}}\ket{\psi_{\cal{A}}}=\sum_k\delta z\,\psi(z_k)\Big(\alpha\ket{+}\ket{z_{k+n}}+\beta\ket{-}\ket{z_{k-n}} \Big)\bigotimes_i\ket{\phi_i,g}\nonumber 
\ee 
is shown to evolve to the correlated one
\be 
\ket{\psi_{\cal{AS}}}=\sum_k\delta z\,\psi(z_k)\Big(\alpha\ket{+}\ket{z_{k+n}}\ket{1_{k+n}}+\beta\ket{-}\ket{z_{k-n}}\ket{1_{k-n}} \Big).
\ee 

We are now in position to introduce to the discussion an element that, although fundamental, is rarely appreciated. It refers to the fact that in every measurement there is at least one degree of freedom that is irremediably discarded, and this is precisely the one about which we want to obtain information. In our example, the fundamentally inaccessible degrees---in fact, that is why we couple an apparatus to get information about them---are the spin and the spatial coordinate of the particle. These degrees of freedom {\em must} be traced out from our theoretical description. This discard is not optional; it is mandatory and irreducible. In doing so we get the following reduced state for the apparatus:
\be \label{rhoA}
\rho_{\cal{A}}=\sum_k\delta z\,|\psi(z_k)|^2\left(|\alpha|^2\ket{1_{k+n}}\bra{1_{k+n}}+|\beta|^2\ket{1_{k-n}}\bra{1_{k-n}} \right).
\ee 
In $\bra{1_i}\rho_{\cal{A}}\ket{1_j}=\delta z\left(|\alpha|^2|\psi(z_{j-n})|^2+|\beta|^2|\psi(z_{j+n})|^2 \right)\delta_{i,j}$ we see that the apparatus state is diagonal in the $\{\ket{1_i}\}$ basis. Then, as far as the observable $\Lambda=\sum_i\lambda_i\ket{1_i}\bra{1_i}$ is concerned, we can ensure via definition \eqref{frakI} that $\mathfrak{I}(\Lambda|\rho_{\cal{A}})=0$, that is, given the available state $\rho_{\cal{A}}$ it follows that $\Lambda$ is real. At the very last stage of the measurement process, information about the apparatus is transported to the observer by photons. In fact, many distinct observers can shine the apparatus and collect their own photons. The point is that the correlations generated between the photons and the apparatus will necessarily be of a classical nature because the state \eqref{rhoA} is an incoherent mixture. Since no quantum correlation is generated and the local irreality of the apparatus (and of the photons) remains null, the relation \eqref{LIQC} guarantees that the reality of the apparatus is preserved during this process. This shows how many observers can get information and agree about the same already-established reality, which thus reveals itself as an objective reality. Also, because the joint state of the apparatus-photons system is correlated only classically, one admits, in light of Bell's theorem, that hidden-variable theories consistent with the hypothesis of local realism are admissible as legitimate models to explain these correlations. In particular, a classical-statistical model such as the Liouvillian theory might accomplish the task in terms of deterministic Hamiltonian trajectories in phase space. However, like quantum mechanics, this model would be unable to predict individual outcomes because uncertainty (in this case deriving from subjective ignorance about the initial state of the system) would still be present. In other words, the inherent statistical character of the formalism precludes precise predictions for individual runs. Hence, given the underlying determinism of such a model, the emerging result of any run of the experiment has to be interpreted as mere information updating, rather than some reality collapse. We claim that this should also be the interpretation for the quantum collapse. The quantum formalism is irreducibly statistical because it was drawn to deal with subtle scenarios involving quantum probability amplitudes, which are associated with pure superpositions. In its statistical capacity it can also describe classical-like behaviors, such as \eqref{rhoA}. Just like the Liouvillian formalism, however, quantum mechanics is not able to predict single outcomes and this should be perfectly fine, since this is what we expected from a theory that deals with (both fundamental and subjective) uncertainties. The final acquisition of information by the observer (who cannot include himself in the theory) is then formulated as an abrupt collapse, which should not be viewed as an actual reduction of any physical element of reality.

Another fundamental point that is not often appreciated in discussions about the measurement problem concerns the notion of quantum reference frames (see, e.g., Refs. \cite{angelo11,angelo12} and references therein). In spite of their complexity~\cite{skotiniotis17}, detectors can be minimally modeled in terms of two degrees of freedom: one related to a visible sign (\{excited,ground\}, as we used above, or \{click,ready\}) and another one related to its location in space-time. Actually, the latter defines the very structure of space-time that plays the role of reference frame. In the discussion above we used the state $\ket{\phi_i}$ for the spatial component of the $i$-th detector. Being very sharp in the configuration space, it presumably is very wide in the momentum space, a feature that is not expected for realistic detectors. In fact, because ordinary detectors are rigidly attached to the laboratory, each one needs to simultaneously have well-defined values of position and velocity at every instant of time, for only in this case can we trust the outcomes we read in each run of the experiment and then make sense of the whole statistics observed. Formally, the observer could describe such an essentially classical detector by admitting that it has an (effective) infinite mass, in which case the uncertainty principle $\Delta z\Delta p\geqslant \hbar/2$ would remain valid whereas $\Delta z$ and $\Delta \dot{z}=\Delta p/m$ vanish simultaneously. In this sense, simultaneous elements of reality for position and velocity emerge from such an intrinsic classicality of the apparatus, which comes from the fact that it is rigidly attached to and therefore defines the reference frame. To a certain extent, we can recognize here the Bohr claim about the irreducibly classical nature of the apparatus. This is not to say, however, that the apparatus is absolutely classical in any sense. In fact, an external observer who can detect the motion of the laboratory would ascribe a finite mass to the apparatus and, as consequence, could eventually find it in superposition \cite{angelo11,angelo12}, that is, with no positional element of reality.

Finally, it is opportune to further elaborated on how the notion of a fundamental irreversibility, in an informational sense, emerges in the present context. The external observer $\cal{O}_{\text{ext}}$, before performing any measurement, describes the joint system $\cal{OAS}$ in terms of a closed dynamics which, as such, preserves the total information associated with $\rho_{\cal{OAS}}(t)$, that is, $\Delta I_{\cal{OAS}}=\Delta S_{\cal{OAS}}=0$. In this case, if provided with precise information about $\rho_{\cal{OAS}}(t)$ and about the interactions among the parts, $\cal{O}_{\text{ext}}$ could theoretically reverse the time evolution of the system and thus get to know the initial state of $\cal{OAS}$. We propose to take this as a statement of informational reversibility. If, on the other hand, $\cal{O}_{\text{ext}}$ is given precise information about the interaction between the internal observer $\cal{O}$ and $\cal{AS}$ but has no access to the resulting state of $\cal{O}$ after the interaction (as in an unrevealed measurement protocol), then the initial ignorance $S_i=S(\sigma_{\cal{AS}})$ that $\cal{O}_{\text{ext}}$ has about $\cal{AS}$ evolves to $S_f=S(\Phi_{A}(\sigma_{\cal{AS}}))$, which means that $\Delta S\geqslant 0$ and $\Delta I\leqslant 0$. Clearly, the lack of information about $\cal{O}$'s state (discard) implies an irreversible decrease of information. In fact, if provided with precise information about the final state $\Phi_A(\sigma_{AS})$ and the interactions between $\cal{A}$ and $\cal{S}$, $\cal{O}_{\text{ext}}$ would {\em not} be able to predict the initial state $\sigma_{\cal{AS}}$. With regard to the internal observer $\cal{O}$, who does not include oneself in the physical description, the information is not preserved as well. The initial ignorance that $\cal{O}$ has about the system is given by $\bar{S}_i\equiv S(\sigma_{\cal{AS}})=S_{\cal{S|A}}+S_{\cal{A}}=S_{\cal{A|S}}+S_{\cal{S}}$. After many runs of the experiment, the average ignorance about the system $\cal{AS}$ is given by $\bar{S}_f=\bar{S}_{\cal{S|A}}+\bar{S}_{\cal{A}}$, where $\bar{S}_{\cal{S|A}}=\sum_ap_aS(\sigma_{\cal{S}|a})$ and $\bar{S}_{\cal{A}}=\sum_ap_aS(A_a)=0$. Since we have $\bar{S}_i\geqslant S_{\cal{S}}=S(\sigma_{\cal{S}})$ and, via concavity, $S(\sigma_{\cal{S}})=S(\sum_ap_a\sigma_{\cal{S}|a})\geqslant \sum_ap_aS(\sigma_{\cal{S}|a})$, it follows that $\bar{S}_i\geqslant \bar{S}_f$. Hence, $\Delta \bar{S}\leqslant 0$ and $\Delta \bar{I}\geqslant 0$. Here the collapse implies gain of information, but this is also an irreversible process because $\cal{O}$ does not describe one's interaction with $\cal{AS}$ and therefore cannot reverse the time evolution to obtain information about $\sigma_{\cal{AS}}$. It is instructive to note that the information increase for $\cal{O}$, in contrast with the discard-induced information decrease for $\cal{O}_{\text{ext}}$, derives from the fact that $\cal{O}$ has access to the sequence $\{a\}$ of outcomes for the apparatus. To see this, note that 
\beq 
\Delta S&=&S(\Phi_A(\sigma_{\cal{AS}}))-S(\sigma_{\cal{AS}})\nonumber \\ 
&=&S(\Phi_A(\sigma_{\cal{A}}))+\sum_ap_aS(\sigma_{\cal{S}|a})-S(\sigma_{\cal{AS}})\nonumber \\
&=&H(\{p_a\})+\Delta \bar{S},
\eeq 
where $H(\{p_a\})$ is the Shannon entropy associated with the distribution $p_a$. It follows that $\Delta\bar{I}=H(\{p_a\})+\Delta I$, which proves the point. The take away message here is as follows: It is the inevitable discard of degrees of freedom associated with the internal observer $\cal{O}$, which receive part of the information flow, that yields the fundamental informational irreversibility perceived by this observer. The external one, who deals with a closed system and no discard of information, has at hand a reversible dynamics. 

Notice that throughout this paper we have taken the von Neumann entropy $S$ purely as an ignorance quantifier, as in any informational framework. However, the Landauer erasure principle~\cite{landauer61,bennett82}, which tells us that information has effective thermodynamical implications, along with recent developments in the emerging field of quantum thermodynamics~\cite{hilt11,goold16}, provides substantial license for one to conceptually connect $S$ with the thermodynamical entropy. In this case, the apparently separated notions of informational (ir)reversibility, which we have assessed so far, and the usual one of thermodynamic (ir)reversibility may coalesce into a single concept.

\section{Concluding remarks}
\label{conclusion}

Quantum mechanics teaches us that the classical deterministic notion of an objective reality calls for a critical review. In this work we employ a recently developed measure of reality~\cite{bilobran15} and traditional tools of quantum information theory to get some insight into the issue. Careful experimental inspections of microscopic systems have pointed out that there are many instances where the physical reality seems to be in suspension, that is, physical quantities do not have well-defined values. As we have shown here, this can be achieved, e.g., by letting a particle interact with massive structures, for in such cases the (apparent frustration of) conservation laws prevent the generation of entanglement and enhance the irreality of a particle's degrees of freedom. Irreality can also be created for a given observable by means of revealed measurements of an incompatible observable. On the other hand, we also showed that any attempt to probe nature, even via arbitrarily tiny monitorings (unrevealed collapse or entanglement plus discard), leads to the emergence of elements of reality. As formally stated in the complementarity relation \eqref{complR}, the flow of quantum information from the system to the apparatus increases the reality of the monitored observable. In a detailed account of the measurement process, we find another facet of this story: Information associated with the apparatus flows to the degree of freedom that we want to measure, the one that is invariably discarded. It follows that the degrees of freedom of the apparatus, in particular those that define the very space-time structure of the reference frame, become real. At this stage, quantum mechanics predicts a fully incoherent mixture for the apparatus, meaning that only subjective ignorance persists about an already established reality. The final (irreversible) flow of information, which is mediated by photons that inform the observer about the state of the apparatus, materialize the information updating of the observer, a step that is out of reach of any statistical theory. In quantum mechanics, this dynamically indescribable transition is called collapse. 

To conclude, it is worth emphasizing that the adoption of BA's notion of reality allows us to formalize a complementarity relation between reality and information. We find in this framework that quantum mechanics predicts no objective reality for isolated systems. Elements of reality can emerge for a given observable only through the codification of information about this quantity. This process, however, does not demand the existence of a brain-endowed system to collect and interpret the information. All that is fundamentally necessary is the presence of physical degrees of freedom that can get correlated with the observable and thus encode information about it. The information that flows to these degrees of freedom makes the reality emerge and become potentially accessible to brain-endowed observers.

\section*{Acknowledgments} 
P.R.D. and R.M.A. acknowledge financial support from CAPES (Brazil) and the National Institute for Science and Technology of Quantum Information (CNPq, Brazil) respectively.


\end{document}